\documentclass[12pt, preprint]{aastex}
\def\deg{\ifmmode^\circ\else$^\circ$\fi}

\def\gs{{_>\atop^{\sim}}}

\def\lsun{L$_{\odot}$}

\def\arcs{\ifmmode {''}\else $''$\fi}
\def\arcm{\ifmmode {'}\else $'$\fi}
\def\parcs{\sa=.07em \sb=.03em
     \ifmmode $\rlap{.}$^{\scriptscriptstyle\prime\kern -\sb\prime}$\kern -\sa$
     \else \rlap{.}$^{\scriptscriptstyle\prime\kern -\sb\prime}$\kern -\sa\fi}
\def\parcm{\sa=.08em \sb=.03em
     \ifmmode $\rlap{.}\kern\sa$^{\scriptscriptstyle\prime}$\kern-\sb$
     \else \rlap{.}\kern\sa$^{\scriptscriptstyle\prime}$\kern-\sb\fi}

\def\Msun{M$_{\odot}$}

\def\spose#1{\hbox to 0pt{#1\hss}}
\def\simlt{\mathrel{\spose{\lower 3pt\hbox{$\mathchar"218$}}
     \raise 2.0pt\hbox{$\mathchar"13C$}}}
\def\simgt{\mathrel{\spose{\lower 3pt\hbox{$\mathchar"218$}}
     \raise 2.0pt\hbox{$\mathchar"13E$}}}
\def\lsim{\rlap{$<$}{\lower 1.0ex\hbox{$\sim$}}}
\def\gsim{\rlap{$>$}{\lower 1.0ex\hbox{$\sim$}}}

\begin{document}

\title{Detection of the Buried AGN in NGC 6240 with the Infrared Spectrograph on The Spitzer Space 
Telescope\altaffilmark{1}}

\author{L. Armus\altaffilmark{2}, 
J. Bernard-Salas\altaffilmark{3},
H.W.W. Spoon\altaffilmark{3},
J.A. Marshall\altaffilmark{3},
V. Charmandaris\altaffilmark{4,7},
S.J.U. Higdon\altaffilmark{3},
V. Desai\altaffilmark{5},
L. Hao\altaffilmark{3},
H.I. Teplitz\altaffilmark{2},
D. Devost\altaffilmark{3},
B.R. Brandl\altaffilmark{6},
B.T. Soifer\altaffilmark{2,5},
J.R. Houck\altaffilmark{3}}

\altaffiltext{1}{based on observations obtained with the Spitzer Space Telescope, which is
operated by the Jet Propulsion Laboratory, California Institute of Technology, under NASA 
contract 1407}
\altaffiltext{2}{Spitzer Science Center, MS 220-6, Caltech, Pasadena, CA 91125}
\altaffiltext{3}{Cornell University, Ithaca, NY 14853}
\altaffiltext{4}{University of Crete, Department of Physics, P.O. Box 2208 GR-71003, Heraklion, Greece}
\altaffiltext{5}{California Institute of Technology, Pasadena, CA 91125}
\altaffiltext{6}{Leiden University, P.O. Box 9513, 2300 RA Leiden, The Netherlands}
\altaffiltext{7}{Chercheur Associ\'e, Observatoire de Paris, F-75014, Paris, France}

\begin{abstract}

We present mid-infrared spectra of the nearby, Ultraluminous Infrared
Galaxy NGC 6240 taken with the IRS on Spitzer. The spectrum of NGC 6240 is
dominated by strong fine-structure lines, rotational H$_{2}$ lines, and PAH
emission features.  The H$_{2}$ line fluxes suggest molecular 
gas at a variety of temperatures.
A simple two-temperature fit to the 
S(0) through S(7) lines implies a mass of 
$\sim 6.7\times 10^{6}$\Msun at T$\sim957$K and
$\sim 1.6\times 10^{9}$\Msun at T$\sim164$K, or about $15\%$
of the total molecular gas mass in this system.  Noteably, we
have detected the [NeV] $14.3\mu$m emission line, with a flux of $5\times
10^{-14}$ erg cm$^{-2}$ s$^{-1}$, providing the first direct detection of
the buried AGN in the mid-infrared.  Modelling of the total SED from near
to far-infrared wavelengths requires the presence of a hot dust
(T$\sim700$K) component, which we also associate with the buried AGN.  The
small [NeV]/[NeII] and [NeV]/IR flux ratios, the relative
fraction of hot dust emission, as well as the large $6.2\mu$m PAH EQW, are
all consistent with an apparent AGN contribution of only $3-5\%$ to the bolometric
luminosity.  However, correcting the measured [NeV] flux by
the extinction implied by the silicate optical depth and our SED fitting,
suggests an intrinsic fractional AGN contribution to the bolometric
luminosity of $\sim20-24\%$ in NGC 6240, which lies within the range
implied by fits to the hard X-ray spectrum.

\end{abstract}


\section{Introduction}

Ultraluminous Infrared Galaxies (ULIRGs), i.e. those galaxies with infrared
luminosity L$_{IR} \gs 10^{12}$\lsun, have the power output of quasars yet emit
nearly all of their energy in the mid and far-infrared part of the spectrum.
Most ULIRGs are found in interacting and merging systems (e.g. Armus, Heckman \&
Miley 1987; Sanders, et al. 1988; Murphy, et al. 1996), where the merger has
driven gas and dust towards the remnant nucleus, fueling a massive starburst,
and either creating or fueling a nascent AGN (Mihos \& Hernquist 1996).

Spectroscopic observations with the ISO satellite greatly expanded our
understanding of the mid-infrared spectra of ULIRGs (e.g., Genzel et al. 1998;
Lutz et al. 1999; Rigopoulou et al.  1999; Spoon et al. 2002; Sturm et al. 2002;
Tran et al. 2001).  However, the the limitations in sensitivity of the ISO
spectrometers, left many ULIRGs, even at relatively low redshift, beyond the
reach of these methods until now.

In order to adequately sample the local ULIRG population, we are obtaining
mid-infrared spectra of a large number of ULIRGs ($ \sim 100$) having $0.02 < z <
0.93$ with the Infrared Spectrograph\altaffilmark{8} (IRS) on Spitzer.  The
entire sample will be observed with the low-resolution ($ \sim 80$) IRS modules,
and about half will be observed with the high-resolution ($ R \sim 650$) IRS
modules as well.  These sources are chosen primarily from the IRAS 1-Jy (Kim \&
Sanders 1998), 2-Jy (Strauss et al. 1992), and the FIRST/IRAS radio-far-IR
sample of Stanford et al. (2000).  In this letter, we present the IRS spectra of
NGC 6240, a nearby ($z=0.0245$), double-nucleus, merging galaxy (Fosbury \& Wall
1979; Fried \& Schulz 1983; Wright, Joseph \& Miekle 1984), with an $8-1000\mu$m
luminosity (L$_{IR}$ as defined by Sanders et al 1988) of $\sim 7\times
10^{11}$\lsun.  The optical nuclear spectrum of NGC 6240 is classified as a
LINER (Armus, Heckman \& Miley 1989), and the extended optical nebula reveals
the presence of a starburst-driven superwind (Heckman, Armus \& Miley 1987;
1990).  X-ray observations with ASCA (Turner et al. 1997), Beppo-SAX (Vignati et
al. 1999), Chandra (Komossa et al. 2003, Ptak et al. 2003), XMM-Newton (Netzer
et al. 2005) provide clear evidence for the presence of one (or two) AGN behind
significant columns of absorbing material (N$_{H} = 1-2\times 10^{24}$
cm$^{-2}$).

Strong [OIV] $25.89\mu$m line emission in the ISO SWS spectrum of NGC
6240 led Lutz et al. (2003) to suggest that up to $50\%$ of the infrared energy
emitted by NGC 6240 could be powered by a buried AGN, yet direct evidence for
the AGN in the mid-infrared was lacking until now.  In this paper we present IRS
spectra that show the presence of faint [NeV] $14.3\mu$m emission from NGC 6240,
and we use fine-structure line ratios and PAH emission feature strengths to
estimate the energetic importance of the AGN.  Because it is a nearby, merging
system that has a powerful starburst, a buried (pair of) AGN and a superwind,
NGC 6240 provides us with a close-up view of the critical processes that shape
all ULIRGs.  Our analysis of the IRS spectra of NGC 6240 will serve to highlight
the advantages, and the limitations, of IRS spectroscopy for studying luminous,
infrared galaxies and buried AGN.

Throughout the paper, we will adopt a flat, $\Lambda$-dominated Universe ($H_0 =
70$ km s$^{-1}$\ Mpc$^{-1}$, $\Omega_M=0.3$, $\Omega_{\Lambda}=0.7$).  The
luminosity distances to NGC 6240 is then 99.2 Mpc, assuming $z=0.0245$,  and one
arcsec subtends $\sim450$ pc in projection.

\section{Observations}

NGC 6240 was observed with the IRS on 04 March 2004.  The IRS is fully described
in Houck et al. (2004).  All four IRS modules were used in Staring mode.  Two
cycles of 14\,sec each were obtained for each nod position and each order of the
Short-Low (SL) and Long-Low (LL) modules.  Six cycles of 30\,sec each, and four
cycles of 60sec each, were obtained for each nod position of the Short-High (SH)
and Long-High (LH) modules, respectively.  A high-accuracy blue peak-up was
performed on a nearby offset star before placing the slits on NGC 6240.  Since
the separation of the twin nuclei in NGC 6240 is about 1.5 arcsec, they are both
contained within all IRS apertures, the narrowest of which (SL) has a slit width
of 3.6 arcsec.

\section{Data Reduction and Analysis}

All spectra were reduced using the S11 IRS pipeline at the Spitzer Science
Center.  This reduction includes ramp fitting, dark sky subtraction, droop
correction, linearity correction, and wavelength and flux calibration.
One-dimensional spectra were extracted from the two-dimensional spectra using
the SMART data reduction package (Higdon et al. 2004).  The one-dimensional
spectra are then scaled to a spectrum of $\alpha$ Lacertae, extracted in an
identical manner using SMART.  The SL and LL data have been sky subtracted by
differencing the two nod positions along the slit, before spectral extraction.
As a final step, we have normalized the SL and LL 1D spectra upwards to match
the IRAS $25\mu$m FSC data (Moshir et al. 1990).  The scale factor was 1.16.
The low-resolution IRS spectra are displayed in Fig. 1a.

Since the SH and LH slits are too small for on-slit background subtraction, we
have subtracted the expected background flux through each slit based on the
model of Reach et al. (2004).  The LH data went through an additional cleaning
step to remove residual hot pixels, which uses the B-mask (supplied with the BCD
data) and the warm pixel mask for the given IRS campaign.  Hot pixels were
interpolated over spatially by using the two nod positions, before extraction
from 2D to 1D spectra.  For a full description of this method, see Devost et al.
(2005).  The SH and LH spectra were then scaled to the corresponding IRS
low-resolution spectra, using a single scale factor for each module.  The scale
factors were 1.15 and 1.07 for SH and LH, respectively.  The high-resolution IRS
spectra are shown in Fig. 1b,c.

\section{Results}

\subsection{Continuum \& Dust Features}

The IRS low-res spectrum of NGC 6240 shows a steeply rising continuum, heavily
absorbed at $10$ \& $18\mu$m by amorphous silicates (Fig. 1a).  There is little
or no water ice or hydrocarbon absorption from $5-7.5\mu$m, as is seen in some
nearby ULIRGs -- e.g. UGC 5101 (Armus et al. 2004).  Broad PAH (Polycyclic
Aromatic Hydrocarbon) emission features at 6.2, 7.7, 11.3, and $12.6\mu$m are
visible.  In addition, we detect the much fainter 16.4, and $17.4\mu$m PAH
features, along with the broad ``plateau" between $16-18\mu$m, which are seen in
some ULIRGs and nearby starburst galaxies (Sturm et al. 2000, Armus et al. 2004,
Smith et al. 2004, Brandl et al. 2005).  The very weak UIB at $14.22\mu$m (see
section 4.2) may also be produced by poly-cyclic aromatic hydrocarbons.


To estimate the contribution of dust emission at different temperatures to the
luminosity of NGC 6240, and to accurately measure the strengths of the PAH
emission features against the underlying, silicate-absorbed continuum, we have
fit the SED with a multi-component model which includes at least two graphite
and silicate dust grain components, PAH emission features (fit with Drude
profiles), a 3500K blackbody stellar component, and unresolved Gaussian emission
lines (to fit the fine structure and H$_{2}$ lines).  To extend the SED, we have
added near-infrared, far-infrared, and sub-mm data to the IRS low-resolution
spectra.  The near-infrared photometry is from Scoville et al. (2000), the
$60\mu$m and $100\mu$m data are from the IRAS Faint Source Catalog (Moshir et
al. 1990), the $120\mu$m, $150\mu$m, $200\mu$m, $450\mu$ and $850\mu$m data are
from Klaas et al. (2001), and the $350\mu$m data are from Benford (1999).  Our
fit to the NGC 6240 SED is shown in Fig. 2.   For a full description of this
model fitting, see Marshall et al. (2005).

Three dust components are required to fit the NGC 6240 continuum spectrum.
The characteristic temperatures of the three components are $27.1 \pm 0.3$K
(cold), $81.4 \pm 1.8$K (warm), and $680 \pm 12$K (hot).  Each component is
composed of graphite and silicate grains (with optical properties from
Draine \& Lee 1984), distributed in size according to the $R_{V} =3.1$ model of
Weingartner \& Draine (2001).
The hot and warm components are modeled as spherically symmetric dust
shells which are optically thin in the mid-infrared.  The dust grains are
distributed with uniform density around the central source, out to a
distance of ten radial scale factors (where the unit radius is taken to be
the distance at which the equilibrium temperature of the grains is equal to
the component temperature). The temperatures of all grains are calculated
as a function of their size, composition, and distance from the
illuminating source. Grain sublimation is taken into account, so that the
size-distribution can vary with radial distance from the central source,
depending upon the temperature of the grains.  The cold component is
modeled as a distribution of grains immersed in a single radiation field
(i.e., they are not radially distributed from a central source). As such,
the grains in this component have a range of temperatures due to their
varying composition and sizes, but not due to their spatial distribution.
It is presumed that the cold dust is responsible for the absorption of
photons from the hot and warm components. Nonetheless, the emission from
the cold grains is treated as optically thin since they are sufficiently
cold to radiate predominantly at far-infrared wavelengths, where their
opacity is extremely small.  Note that the range of grain temperatures due
to the distribution of grain-sizes in the cold component produces an SED
that is wider than that obtained with an average grain gray-body of the
same temperature.  In this treatment, there is therefore not a unique
temperature associated with each component, as there is in a gray-body
model. We instead define a ``characteristic" temperature of a component to
be the temperature of the most luminous grain size at the distance from the
source contributing the majority of the luminosity. This luminosity
dominating distance corresponds to a $\tau_{UV} \sim 0.5$, where
approximately half of the UV-source photons have been absorbed. Each
component therefore contains dust above and below the characteristic
temperature. With this definition, the characteristic temperature roughly
corresponds to the expected peak in the dust modified Planck function.

A hot dust component with a characteristic temperature of about 700K, as we have
found in NGC 6240, implies the existence of a significant number of grains near sublimation.
As suggested by Laurent et al. (2000), an excess of
hot dust can be taken as evidence for grains heated close to sublimation in the
presence of an AGN.  In NGC 6240, the presence of this hot dust together with
the high-ionization lines discussed below, are consistent with this
interpretation.  The hottest grains in thermal equilibrium in the T$=680$K dust
component, if illuminated with an AGN-like spectrum (Sanders et al. 1989) having
the bolometric luminosity of NGC 6240, would be at a distance of about 1 pc from
the central source.  The hot component accounts for approximately $3.5\%$ of the
bolometric luminosity in NGC 6240.

The optical depths to both the hot and cold dust components are essentially
zero, while the screen optical depth to the warm component is
$\tau_{W}=5.2\pm0.1$ (effectively the silicate optical depth, since this
component dominates at $9.7\mu$m), implying an A$_{V}\sim 95\pm 2$ mag.
This is about a factor of three larger than the lower limit which would be
derived from a smooth continuum fit anchored at $5.3-5.6\mu$m, $14\mu$m and
$34\mu$m, since the un-extinguished model continuum at $9.7\mu$m is
dominated by silicate emission.  The warm component accounts for about
$26\%$ of the bolometric luminosity before correcting for extinction, and
about $80\%$ after correcting for this extinction.

While a multi-component SED fit is never ``unique", we have compared our best
fit to alternative models which exclude a hot dust component, and which tie
the optical depth to the hot dust at the value derived for 
the warm component. The latter is most similar to a spherical shell model
wherein the hot dust is closest to the (obscured) nucleus and the coldest
dust is responsible for most of the obscuration along the line of sight.
Both of these constrained fits produce significantly poorer results, 
producing reduced chi-squared values that are between 50\% and a factor
of two larger than our preferred, three component fit described above.

Fits to the multiple PAH emission features in NGC 6240 result in 6.2/7.7
and 11.3/7.7 flux ratios of 0.21 and 0.25, respectively, suggesting a mix
of neutral and ionized grains with between $200-400$ Carbon atoms each
(Draine \& Li 2001).  The 6.2/7.7 flux ratio is between that found for NGC
253 and the Circinus galaxy, whereas the 11.3/7.7 flux ratio is about a
factor of two, larger (Sturm et al.  2000; Draine \& Li).  The $6.2\mu$m
PAH EQW is 0.52, about $20-30\%$ smaller than what is found for pure
starburst galaxies (Brandl et al.  2005), but much larger (a factor of
$5-10$) than is found for AGN.  The PAH-to-L$_{IR}$ lumninosity ratio is
about 0.023.  In all cases, the extinction to the PAH features is assumed
to be zero, as in a model where the emission arises from an extended region
seen against a (partially) obscured continuum.  Note, our multiple
component continuum fits produce significantly larger PAH flux in features
(e.g. the 7.7, 8.6, and 11.3$\mu$m features) that are affected most by
silicate absorptions than do other methods which employ spline fits to the
observed continuum, even those anchored at relatively ``clean" parts of the
spectrum. However, the fitting we have used to derive the PAH fluxes in
Table 1 are directly comparable to the Draine \& Li (2001) models.  This
will be discussed fully in Marshall et al. (2005).

\subsection{Emission Lines}

The IRS short-high and long-high spectra of NGC 6240 are dominated by
unresolved atomic, fine-structure lines of Ne, O, Si, S, and Fe,
covering a large range in ionization potential, (Figs. 1b,1c and Table 1).
The [FeII] lines at 5.34 and $25.99\mu$m (partially blended with 
[OIV] $25.89\mu$m) are particularly strong in NGC 6240 as compared to 
other ULIRGs and starburst galaxies, as has been noted by Lutz et al. (2003).
We also detect a faint feature at $\sim17.9\mu$m which we attribute
to [FeII] emission.  Grain destruction, possibly in the starburst-driven
superwind, is the likely source of the increased gas phase Fe
abundance in the cooling ISM in NGC 6240 
(Draine \& Woods 1990).

The most striking feature of the NGC 6240 spectrum is the strength of the
pure rotational emission lines of H$_{2}$.  As noted by Draine \& Woods
(1990), based primarily upon the strength of the near-infrared H$_{2}$
lines, NGC 6240 has both the largest H$_{2}$ line luminosity, and the
largest ratio of H$_{2}$ -to- infrared luminosity, of any known galaxy. 
In the IRS spectrum, we detect the S(0) through S(7) lines (see Table 1), except for 
S(6), which is blended with the $6.2\mu$m PAH emission feature.
The line flux we measure for the S(0) line is $8.6 (\pm 2.4) \times
10^{-14}$ erg cm$^{-2}$ s$^{-1}$, consistent with the upper limit ($ <
22\times 10^{-14}$ erg cm$^{-2}$ s$^{-1}$) in Lutz et al. (2003). 
The H$_{2}$ excitation level diagram (Fig. 3) provides strong evidence
for multiple temperature components in NGC 6240, a fact noted previously
by Lutz et al.  Fitting the S(3) through S(7) features, we derive a gas
temperature of T$\sim957$K, and a mass of $9.5\times 10^{6}$\Msun~ for 
this warm component.  Subtracting this fit off the lower excitation lines,
we can then fit the S(0) and S(1) lines and derive a gas temperature of 
T$\sim164$K, and a corresponding mass of $1.6\times 10^{9}$\Msun.  In all
cases an ortho--to--para ratio of three is assumed.  Obviously the coldest
component dominates the mass of ``warm" H$_{2}$ in NGC 6240.


A mass of $1.6\times 10^{9}$\Msun~ is about $15\%$ of the molecular
gas mass derived from single-dish millimeter CO line measurements by
Solomon et al. (1997).  However,it is about $50\%$ of the cold
molecular gas, or about 1/3 the total (warm plus cold) molecular gas mass, 
within the central 1 kpc as measured 
by Tacconi et al.  (1999), which itself about 
half of the dynamical mass within the central 1 Kpc. 
The cold gas, as well
as the warmer gas traced by the near-infrared $1-0$ S(1) line (Tecza
et al. 2000), is actually centered between the two nuclei in NGC 6240,
the latter presumably thermally excited via slow shocks triggered by
either the interaction or the outflowing wind (Tecza et al., Van der
Werf 1996, Egami et al.  1998).  Although we have no spatial
information on the H$_{2}$ emission in our IRS spectrum, the
de-convolved linewidths are comparable to those measured by Tecza et
al. ($500-600$km s$^{-1}$), suggesting that much of the warm molecular
gas we measure is associated with the centrally-concentrated gas seen
by Tacconi et al.  and Tecza et al. between the two nuclei.

The strongest H$_{2}$ line in our IRS spectrum, the $0-0$ S(5)
transition at $6.9\mu$m, has a flux about a factor of $4-5$ times
larger that of the near-infrared $1-0$ S(1) line (Rieke et al. 1985;
Herbst et al. 1990; Fischer, Smith \& Glaccum 1990).  The mid-infrared
H$_{2}$ emission lines listed in Table 1 alone account for about
$0.16\%$ of the total energy output of NGC 6240.  The model proposed
by Draine \& Woods (1990) for the strong H$_{2}$ emission in NGC 6240, that
of thermal emission from gas heated by X-rays, also predicts an $16.33\mu$m
H$_{3}^{+}$ rotational feature with a flux of $\sim 2.5\times
10^{-14}$erg cm$^{-2}$ s$^{-1}$.  This feature is unique to the X-ray
heating model.  Unfortunately, this key diagnostic falls near the
broad, PAH emission feature at $16.4\mu$m, which is clearly seen in
our IRS spectrum.  Since our limits for unresolved features in this
part of the spectrum are about $3\times 10^{-14}$erg cm$^{-2}$
s$^{-1}$, even without the presence of the PAH emission, we cannot
confirm or rule out the presence of the  H$_{3}^{+}$ line at the level
predicted by Draine \& Woods.

Most of the emission lines are resolved in our SH and LH spectra, with the
bright lines having an average linewidth of $\sim 680$ km s$^{-1}$,
corrected for instrumental broadening, 
comparable to the linewidths reported by Lutz et al. (2003) from their SWS
spectra.  There is no obvious trend of increased linewidth with wavelength.

Ratios of the mid-infrared atomic fine structure emission lines can be
used to characterize the source of the UV photons which ionize the gas
and heat the dust (Genzel et al. 1998, Lutz et al. 1998, Sturm et al.
2002, Verma et al.  2003).  Some features, e.g. the [NeV] lines at
14.3 and $24.3\mu$m, imply the presence of an AGN, since the
ionization potential (97.1 eV) is too large to be produced by main
sequence stars.  The same is not true for [OIV], since it takes only
55 eV to ionize O$^{++}$. While the $25.89\mu$m is strong in the spectra
of AGN, it is seen in many starburst galaxies as well (e.g., Lutz et
al.  1998, Verma et al. 2003).  In the IRS spectrum of NGC 6240, [OIV]
has a line flux of $27.1\times 10^{-21}$W cm$^{-2}$, and the [OIV]
25.9/[NeII] 12.8 line flux ratio is 0.12.  The [SIII] 18.7/[SIII] 33.4
line flux ratio (0.75) suggests an electron density of n$_{e}\sim
400$cm$^{-3}$ in the ionized nebula, for T$=10^{4}$K.

An expanded view of the rest-frame $13.7-15.0\mu$m region in NGC 6240 (order 14
from the SH spectrum) is shown in Fig. 4a.  Two broad features are evident -- a
weak feature roughly centered at $14.2\mu$m, and a second, stronger feature,
roughly centered at $14.35\mu$m.  We identify the first as $14.22\mu$m PAH
emission (Moutou et al. 1996), and the second as a blend of the [NeV]
$14.32\mu$m and [ClII] $14.36\mu$m fine structure emission lines.  The [NeV]
line is a direct indication of the AGN nature of NGC 6240. The [ClII] line has
been seen in the mid-infrared spectra of a number of starburst galaxies (Sturm
et al 2000, Spoon et al. 2000, Devost et al. 2005).  The broad, [NeV], [ClII]
blend is not seen in other ULIRGs in our sample with [NeV] emission (see Fig.4b
and Armus et al. 2005), and is not an instrumental artifact.  In Fig. 4b we
have overlayed the same part of the spectrum from Mrk 273, another nearby ULIRG
with a strong [NeV] emission line (Armus et al.  2005).  In Mrk 273 we see
neither the $14.22\mu$m PAH nor the $14.36\mu$m [ClII] line.  In NGC 6240 we
have fit the [NeV] + [ClII] complex with two Gaussians having fixed positions,
and widths equal to that measured for the isolated [NeIII] $15.55\mu$m line
(Fig. 4c).  Only the relative flux in the lines is allowed to vary.  The [NeV]
$14.32\mu$m emission line flux is then $5.1\times 10^{-21}$ W cm$^{-2}$.  After
removal of the [NeV] and [ClII] lines, the broad feature at $14.2\mu$m is fit
with an single, un-constrained Gaussian.  The [NeV] 14.3/[NeII] 12.8, and the
[NeV] 14.3/[NeIII] 15.5 line flux ratios are then $\sim0.03$ and $\sim0.08$,
respectively.  We have not detected the [NeV] $24.32\mu$m line, with an upper
limit on the line flux of $3.9\times 10^{-21}$W cm$^{-2}$.  The [NeV] 14.3/24.3
line flux ratio is $>1.3$, comparable to that found for nearby, bright Seyfert 2
galaxies by Sturm et al. (2002).

\section{Discussion}

We have detected the [NeV] emission line in the IRS spectrum of NGC 6240.
This is the first reported detection of this feature, and it provides a
direct measure of the buried AGN.  Although NGC 6240 does exhibit
large-scale shocks from the superwind, it is unlikely that these shocks
contribute to the [NeV] emission.  As pointed out by Voit (1992), shock
models tend to enhance the low-ionization lines (e.g., [NeII] or [SiII]).
While our measured [NeII] 12.8/[NeIII] 15.55 line flux ratio of 2.8 is
larger than found in most pure AGN models (Voit 1992), it is much lower
than is typically found in shocks, where [NeII]/[NeIII] $\ge 10$ (Binette,
Dopita \& Tuohy 1985).  It is, however, completely consistent with a
moderate excitation starburst galaxy (Thornley et al. 2000, Verma et al.
2003).  We therefore favor a model wherein the [NeV] emission in NGC 6240
comes from a coronal-line region in close proximity to one or both of the active
nuclei, and that most of the observed [NeII] and [NeIII] emission is dominated
by the surrounding starburst.  

Pure AGN, and AGN-dominated ULIRGs, typically have [NeV]/[NeII] and
[OIV]/[NeII] line flux ratios of $0.8-2$ and $1-5$, respectively (Sturm et
al. 2002, Armus et al. 2004).  Starburst galaxies, on the other hand
typicall have stringent upper limits on [NeV] ([NeV]/[NeII] $< 0.01$), yet
a large range in [OIV]/[NeII] flux ratios, from $0.01 - 0.2$ (Sturm et
al.  2002, Verma et al. 2003, Devost et al.  2005).  Therefore, both the
[NeV]/[NeII] and [OIV]/[NeII] line flux ratios in NGC 6240 are
significantly below the AGN values, while the [OIV]/[NeII] ratio is on the
high end, but still consistent with that found in some high-excitation
starburst galaxies.  If we use a simple linear mixing model (e.g. Sturm et
al. 2002) wherein the observed values of [NeV]/[NeII] and [OIV]/[NeII]
seen in NGC 6240 are due to ``excess" [NeII] from the starburst, then both
line ratios imply an apparent AGN contribution of about $3-5\%$ to the
bolometric luminosity.  The detection of a hot dust component, which is
not seen in IRS spectra of starburst galaxies (Brandl et al. 2005), and
which is responsible for $\sim3-4\%$ of the bolometric luminosity, is
additional indirect evidence for the buried AGN.

Fits to the Beppo-SAX, hard X-ray spectrum (Vignati et al. 1999) suggest
an HI column density toward the active nucleus of $1-2\times 10^{24}$
cm$^{-2}$, or an A$_{V} \sim 500-1000$ mag.  Correcting for this HI column
implies an intrinsic 2-10 KeV, hard X-ray to bolometric luminosity of
$0.7-2\times 10^{44}$ erg s$^{-1}$, or a hard X-ray to bolometric
luminosity ratio of $\sim 0.025-0.075$, depending upon the unknown fraction
of the X-ray flux which is reflected.  This range is comparable to other
AGN which typically have hard X-ray to bolometric flux ratios of $0.01-0.2$
(see Ptak et al. 2003 for a summary), suggesting that NGC 6240 has a rather
``normal" hard X-ray to bolometric luminosity ratio for an AGN (Vignati et
al. 1999).  The extinction-corrected, hard X-ray data are therefore
consistent with the buried AGN producing anywhere from $10-100\%$ of the
luminosity in NGC 6240, with the larger values much more likely.  Note, the
typical (Elvis et al. 1994) QSO luminosity ratio is about 0.03.

While the [NeV]/[NeII] line flux ratio is often treated as reddening
independent, this is not strictly true.  Since the [NeV] is coming from
gas directly heated by the AGN, while the [NeII] can be dominated by the
extended starburst, the distribution of the dust in NGC 6240 (and other
ULIRGs) can have a significant effect on the measured [NeV]/[NeII]
emission-line flux ratio.  This affect would typically be to lower the
observed ratio, causing ULIRGs to look more starburst-like.  Could the low
[NeV] flux in NGC 6240 be the result of the large extinction implied by
the X-ray spectrum ?  Correcting the measured [NeV] emission by an
A$_{V}=1000$ mag would produce an intrinsic [NeV]/IR flux ratio in NGC
6240 well above any other known AGN or ULIRG, so the hard X-ray derived
extinction is not applicable.  From our SED fit, the extinction to the
warm dust is A$_{V}\sim 95$ mag.  While large, this is still likely to be
a lower limit to some lines of sight to the circum-nuclear region directly
excited by the AGN.  However, if we assume A$_{V} =95$ mag to the coronal
line region, and an intrinsic [NeV]/[NeII] line flux ratio of 1.0 for this
AGN-heated gas, we derive a corrected [NeV]/[NeII] ratio of 0.24 (or 0.20 if we also
extinction correct the [NeII] which is excited by the AGN).  This implies
the ``intrinsic" AGN contribution is likely to be between $20-24\%$, a
number which is within the range estimated from the fits to
the extinction-corrected, hard X-ray data.

\acknowledgements

We would like to thank Bruce Draine, Aaron Evans, David Hollenbach, Lisa
Kewley and Bill Reach for many helpful discussions. The comments of an
anonymous referee also helped to improve the content and presentation of
this paper.  Support for this work was provided by NASA through an award
issued by JPL/Caltech.

\begin{figure*}[low]
\plotone{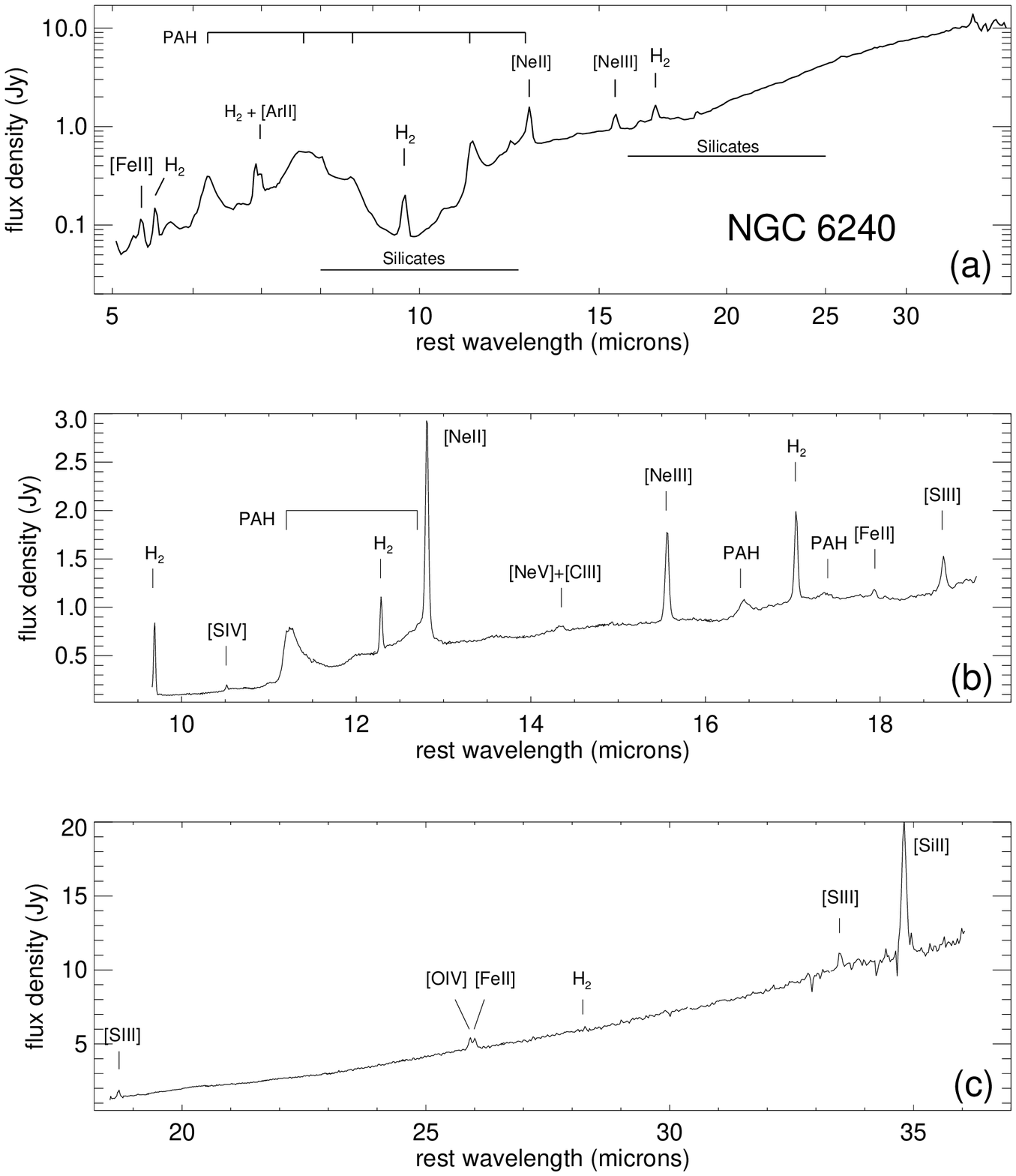}
\caption{IRS Short-Low \& Long-Low (top panel), Short-High (middle panel), and Long-High 
spectra (bottom panel) of NGC 6240.  Prominent 
emission lines and absorption bands (the latter indicated by horizontal bars) are marked.}
\end{figure*}

\begin{figure*}[totalfit]
\plotone{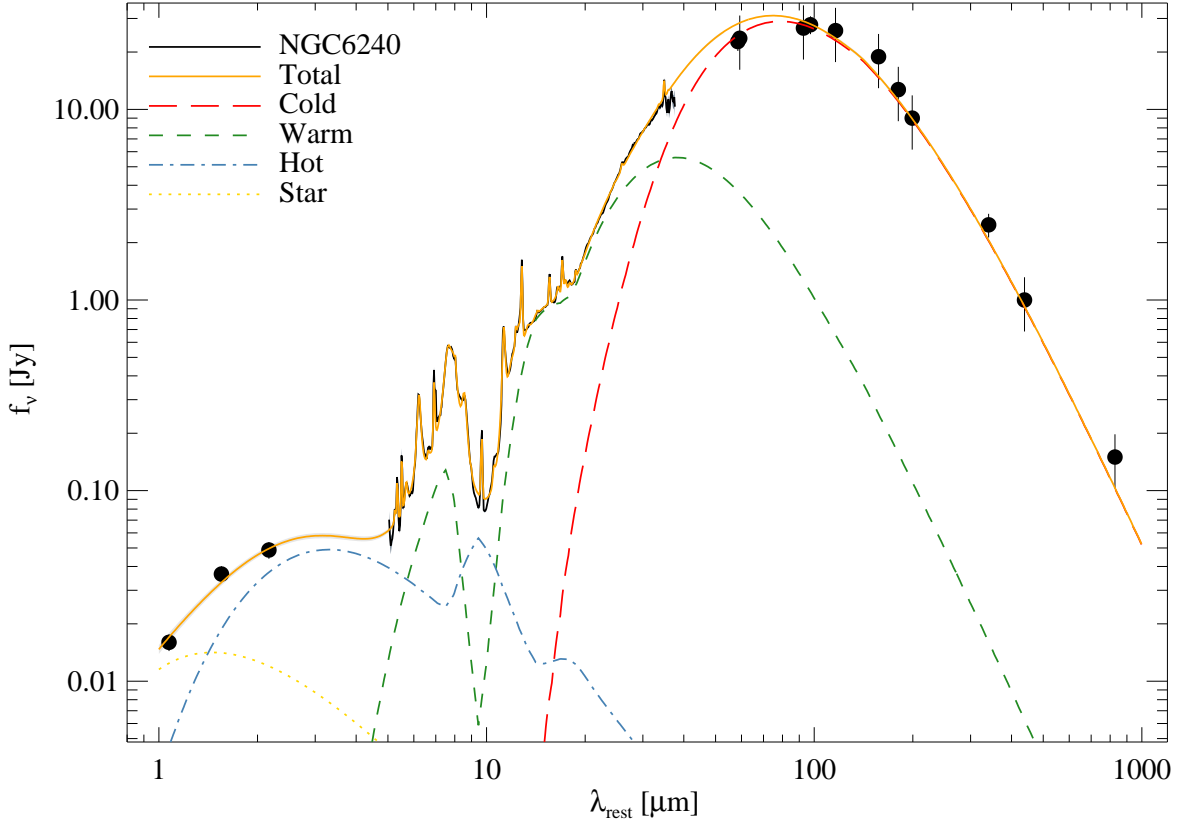}
\caption{Multi-component fit to the NGC 6240 spectrum.  In the first
panel, the fit to the entire SED from $\sim1-1000\mu$m is shown.  In the
second panel only the IRS low-res wavelength range is shown.  In both
panels the orange solid line represents the total fit, and various
spectral components (e.g., stellar light, hot dust, PAH emission, cold
dust) are highlighted individually.  In panel one, only the stellar
emission (yellow dotted line) and the three continuum dust components, hot
(blue, single dot-dashed line), warm (green, short dashed line), and cold (red, long dashed
line) are shown.  In panel two, these four components, along
with the PAH emission (purple, triple dot-dashed line) are also shown.  The
fitting process is described in the text.  Fits to the fine-structure and
H$_{2}$ emission lines have been omitted for clarity.} \end{figure*}

\begin{figure*}[irsfit]
\plotone{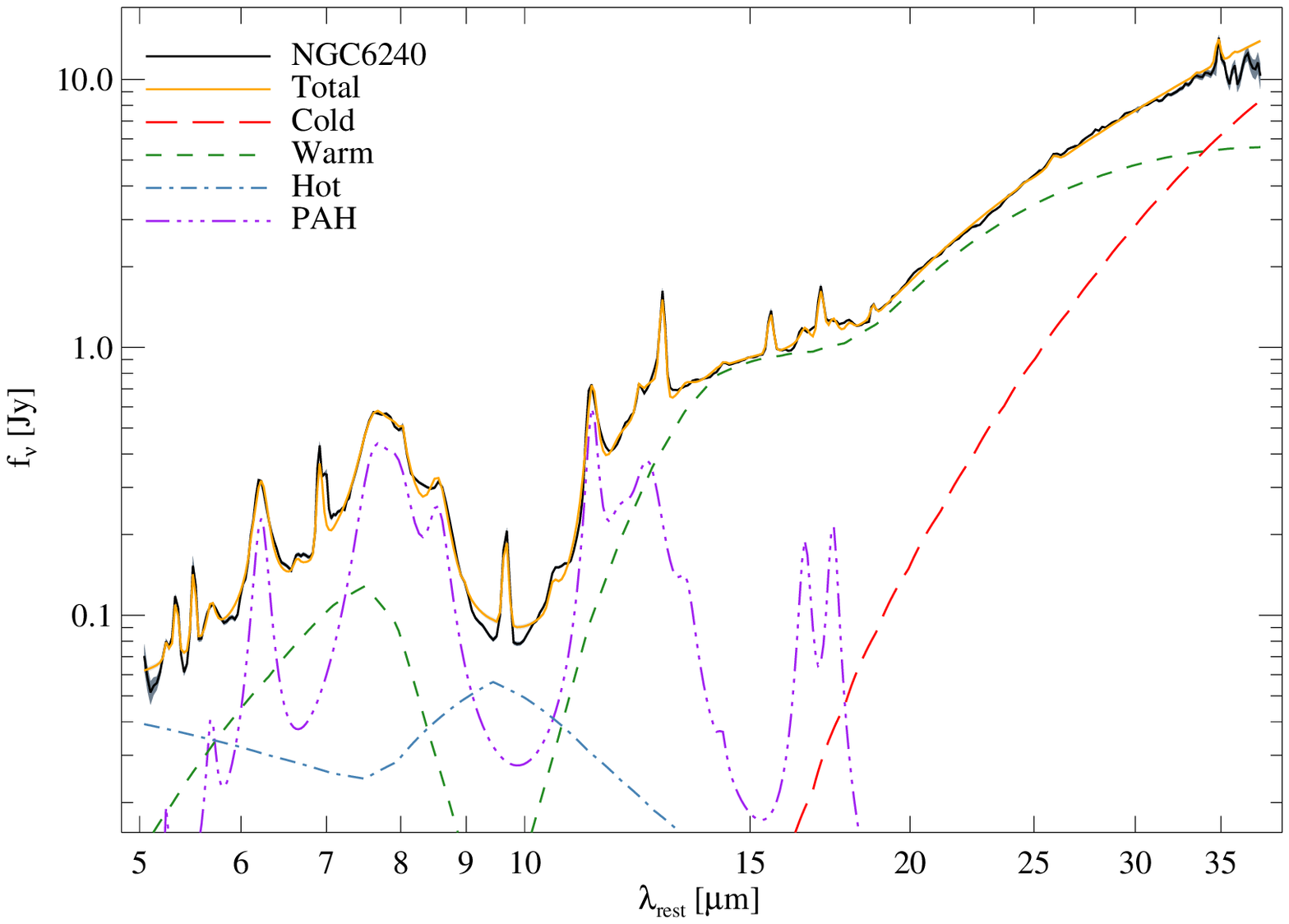}
\end{figure*}

\begin{figure*}[h2]
\plotone{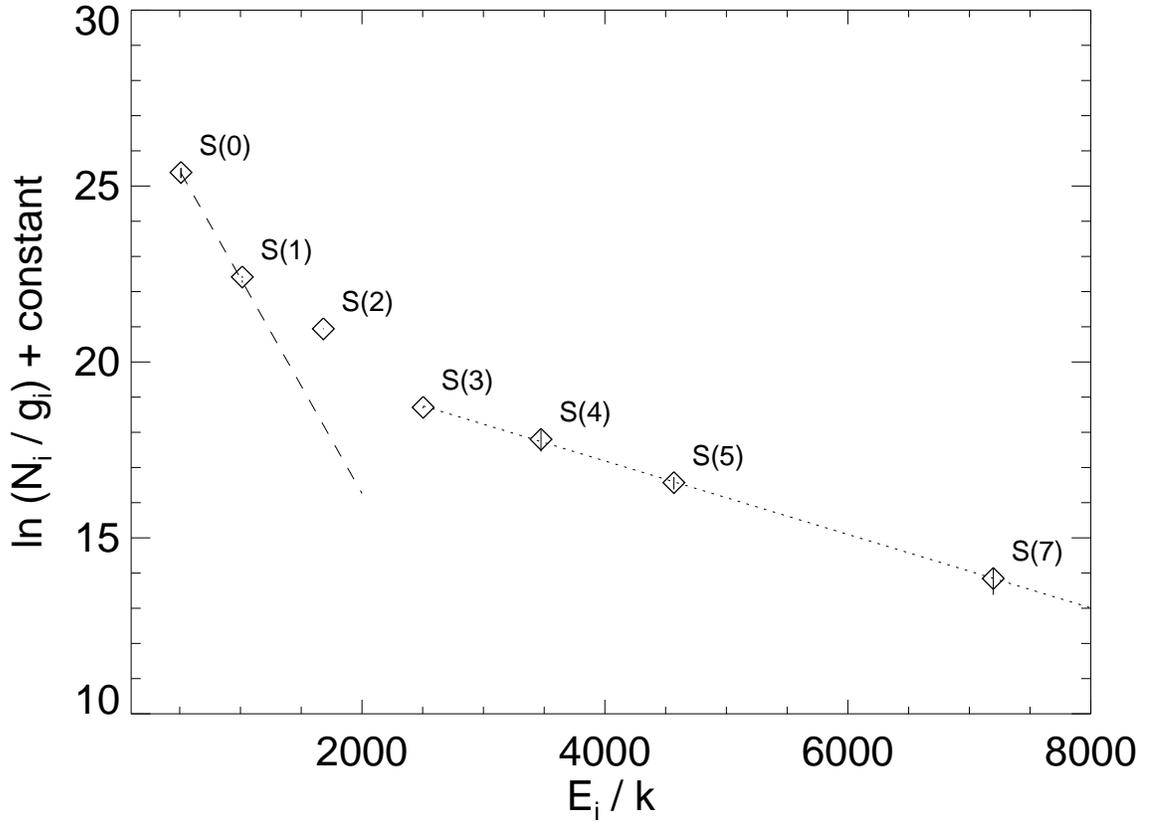}
\caption{H$_{2}$ excitation diagram for NGC 6240, showing for the lines
observed with the IRS, the upper level population divided by the level
degeneracy, as a function of upper level energy.  The lines have not been
corrected for extinction.  Errors are indicated as vertical bars for each
line.  The fits to the hot (T$\sim 957$K) and cold (T$\sim 164$K) components
are indicated by a dotted and a dashed line, respectively.} \end{figure*}

\begin{figure*}[nev] 
\plotone{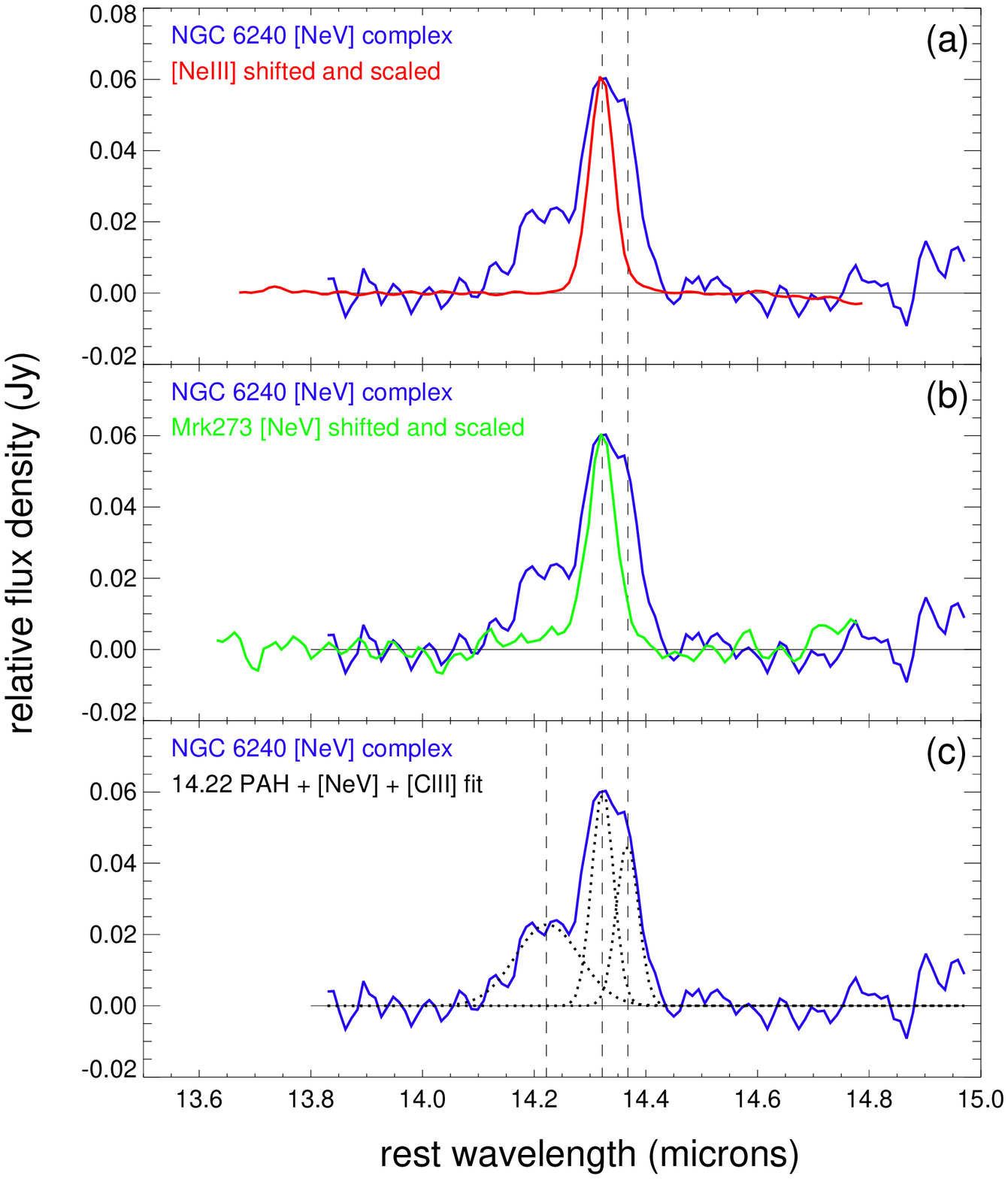} 
\caption{Short-High, order 14
spectrum of NGC 6240, highlighting the PAH, [NeV], [ClII] complex at
$14.2-14.4\mu$m.  In (a) we show the NGC 6240 lines with the [NeIII]
$15.55\mu$m line shifted and scaled to the [NeV] $14.3\mu$m peak.  In (b)
the NGC 6240 features are compared to those seen in another local ULIRG
with [NeV] emission, Mrk 273 (see Armus, et al. 2005). In (c) we show our
fit to the NGC 6240 [NeV] complex with the three Gaussian features in
dashed lines.  In all cases the data have been smoothed with a
three-pixel boxcar, and a linear fit to the continuum has been subtracted
off to place the spectra on a horizontal scale.}
\end{figure*}

\references

\reference{} Armus, L., Heckman, T.M, \& Miley, G.K. 1987, AJ, 94, 831.

\reference{} Armus, L., et al. 2004, ApJ Suppl., 154, 178.

\reference{} Armus, L., et al. 2005, in preparation.

\reference{} Brandl, B.R., et al. 2005, in preparation.

\reference{} Benford, D. 1999, PhD Thesis, California Institute of Technology.

\reference{} Devost, D., et al. 2005, in preparation.

\reference{} Draine, B.T. \& Lee, H.M. 1984, ApJ, 285, 89.

\reference{} Draine, B.T. \& Woods, D.T. 1990, ApJ, 363, 464.

\reference{} Draine, B.T. \& Li, A. 2001, ApJ, 551, 807.

\reference{} Egami, E. 1998, in Proc. IAU Symposium 186, Galaxy Interactions at Low and High
Redshift, ed. J. Barnes \& D.B. Sanders (Dodrecht: Kluwer).

\reference{} Evans, A.S., Mazzarella, J.M., Surace, J.A., \& Sanders, D.B. 2002, ApJ, 580.

\reference{} Fischer, J., Smith, H.A., \& Glaccum, W. 1990, in Astrophysics with Infrared Arrays, 
ed. R. Elston (ASP Conf. Series).

\reference{} Fried, J.W., \& Schulz, H. 1983, A\&A, 118, 166.

\reference{} Fosbury, R.A.E., \& Wall, J.V. 1979, MNRAS, 189, 79.

\reference{} Genzel, R., Lutz, D., Sturm, E., Egami, E., Kunze, D., 
et al. 1998, ApJ, 498, 589.

\reference{} Heckman, T.M., Armus, L., \& Miley, G.K. 1987, AJ, 93, 276.

\reference{} Heckman, T.M., Armus, L., \& Miley, G.K. 1990, ApJ Suppl., 74, 833.

\reference{} Herbst, T.M., et al. 1990, ApJ, 99, 1773.

\reference{} Higdon, S.J.U., et al. 2004, PASP, 116, 975.

\reference{} Houck, J.R., et al. 2004, ApJ Supplement, 154, 18.


\reference{} Imanishi, M., Terashima, Y., Anabuki, N., \& Nakagawa, T. 2003, ApJ, 596, L167.

\reference{} Kim, D.C. \& Sanders, D.B. 1998, ApJS, 119, 41.

\reference{} Klaas, U., et al. 2001, A\&A, 379, 823.

\reference{} Laurent, O., et al. 2000,  A\&A, 359, 887.

\reference{} Lonsdale, C.J., Smith, H.E., \& Lonsdale, C.J. 1995, ApJ, 438, 632.

\reference{} Lutz, D., Kunze, D., Spoon, H.W.W., \& Thornley, M.D. 1998, A\&A, 333, L75.

\reference{} Lutz, D., Veilleux, S., \& Genzel, R. 1999, ApJ, 517, L13.

\reference{} Lutz, D., et al. 2003, A\&A, 409, 867.

\reference{} Marshall, J.A., et al. 2005, in preparation.

\reference{} Mihos, C.J., \& Hernquist, L. 1996, ApJ, 464, 641.

\reference{} Moshir, et al. 1990, IRAS Faint Source Catalog, V2.0.

\reference{} Moutou, C., Verstraete, L., Leger, A., Sellgren, K., \& Schmidt, W. 2000, 
A \& A, 354, L17.

\reference{} Murphy, T.W. Jr., Armus, L., Matthews, K., Soifer, B.T., Mazzarella, J.M., 
Shupe, D.L., Strauss, M.A., \& Neugebauer, G. 1996, AJ, 111, 1025.

\reference{} Ptak, A., et al. 20003, ApJ, 592, 782.

\reference{} Reach, W.T., Morris, P., Boulanger, F., \& Okumura, K. 2004, Icarus, in press.

\reference{} Rieke, G.H. et al. 1985, ApJ, 290, 116.


\reference{} Rigopoulou, Spoon, H.W.W., Genzel, R., Lutz, D., Moorwood, 
 A.F.M., \& Tran, Q.D. 1999, AJ, 118, 2625.

\reference{} Sanders, D.B., et al. 1988 ApJ, 325, 74.
 



\reference{} Smith, J.D.T., et al. 2004, ApJ Supplement, 154, 199.

\reference{} Solomon, P.M., Downes, D., Radford, S.J.E., \& Barrett, J.W. 1997, ApJ, 478, 144.

\reference{} Spoon, H.W.W., Koornneff, J., Moorwood, A.F.M., Lutz, D., Tielens, A.G.G.M. 2000, A\&A, 357, 898.

\reference{} Spoon, H.W.W., Keane, J.V., Tielens, A.G.G.M., Lutz, D., Moorwood, A.F.M., \&
Laurent, O. 2002, A \& A, 385, 1022.

\reference{} Strauss, M.A., Huchra, J.P., Davis, M., Yahil, A., Fisher, K.B., 
\& Tonry, J. 1992, ApJS, 83, 29.

\reference{} Sturm, E., et al. 2000, A \& A, 358, 481.

\reference{} Sturm, E., et al. 2002, A \& A, 393, 821.

\reference{} Tacconi, L.J., Genzel, R., Tecza, M., Gallimore, J.F., Downes, D., \& Scoville, N.Z. 1999, 
ApJ, 524, 732.

\reference{} Tecza, M., Genzel, R., Tacconi, L.J., Anders, S., Tacconi-Garman, L.E., \& 
Thatte, N. 2000, ApJ, 537, 178.

\reference{} Tran, Q.D., et al. 
2001, ApJ, 552, 527.



\reference{} Van der Werf, P.P. 1996, in Cold Gas at High Redshift, ed. M.N. Bremmer, P.P. Van der
Werf, H.J.A. Roettgering, \& C.L. Carilli (Dodrecht: Kluwer).

\reference{} Verma, A., Lutz, D., Sturm, E., Sternberg, A., Genzel, R., \& Vacca, W. 2003, A \& 
A, 403, 829.

\reference{} Vignati, P., et al. 1999, Astron. \& Astrophys., 349, 57.

\reference{} Voit, G.M. 1992, ApJ, 399, 495.

\reference{} Weingartner, J.C. \& Draine, B.T. 2001, ApJ, 548, 296.

\begin{deluxetable}{lcccc}
\tabletypesize{\scriptsize}
\tablecaption{Emission Features\label{tbl-1}}
\tablehead{
\colhead{Feature ID} &
\colhead{$\lambda_{rest}$} &
\colhead{Flux} &
\colhead{EQW} &
\colhead{FWHM} \\
\colhead{} &
\colhead{($\mu$m)} &
\colhead{($10^{-14}$erg cm$^{-2}$s$^{-1}$)}&
\colhead{($\mu$m)}&
\colhead{(km s$^{-1}$)}
}

\startdata
$[FeII]$&5.34&18.9 (6.9)&0.024 (0.009)\\
H$_{2}$ S(7)&5.511&33.7 (12.5)&0.042 (0.015)\\
PAH&6.2&399 (44)&0.52 (0.06)\\
H$_{2}$ S(5)&6.909&95.2 (16.6)&0.095 (0.016)\\
$[ArII]$&6.98&57.0 (15.5)&0.057 (0.015)\\
PAH&7.7&1876 (212)&2.60 (0.28)\\
H$_{2}$ S(4)&8.025&36.4 (10.8)&0.058 (0.017)\\
PAH&8.6&439 (43)&1.35 (0.13)\\
H$_{2}$ S(3)&9.665&70.9 (1.9)&0.2393 (0.0041)&570 (60)\\
$[SIV]$&10.511&3.8 (0.3)&0.0099 (0.0004)&411 (60)\\
PAH&11.3& 475 (31)&1.61 (0.10)\\
H$_{2}$ S(2)&12.279&39.8 (0.5)&0.0334 (0.0007)&542 (60)\\
PAH&12.6&522 (90)&0.78 (0.14)\\
$[NeII]$&12.814&193.1 (3.7)&0.1198 (0.0018)&799 (60)\\
PAH&14.2&5.8 (1.1)&0.005 (0.001)\\
$[NeV]$&14.322&5.1 (0.9)&0.0087 (0.0008)\\
$[ClII]$&14.368&3.6 (1.0)&0.0072 (0.0009)\\
$[NeIII]$&15.555&70.4 (2.4)&0.0625 (0.0019)&828 (60)\\
PAH&16.4&22.4 (0.8)&0.0198 (0.0008)\\
H$_{2}$ S(1)&17.03&50.4 (1.5)&0.0423 (0.0017)&585 (60)\\
PAH&17.4&7.3 (1.0)&0.0055 (0.0010)\\
$[FeII]$&17.934&4.6 (0.7)&0.0039 (0.0006)&657 (62)\\
$[SIII]$&18.713&19.9 (1.4)&0.0178 (0.0022)&810 (120)\\
$[NeV]$&24.318&$< 3.9$&$< 0.0021$\\
$[OIV]$&25.890&27.2 (0.7)&0.0234 (0.0006)&674 (60)\\
$[FeII]$&25.988&23.1 (0.1)&0.0210 (0.0003)&624 (60)\\
H$_{2}$ S(0)&28.218&8.6 (1.2)&0.0032 (0.0004)&439 (90)\\
$[SIII]$&33.481&26.3 (8.4)&0.0093 (0.0030)&537 (60)\\
$[SiII]$&34.815&268.9 (10.1)&0.0981 (0.0005)&823 (60)\\
\enddata

\tablecomments{Line fluxes and rest-frame equivalent widths for
lines as measured in the IRS spectra.  Uncertainties in the line
fits (all single Gaussians except where noted) are listed in
parentheses.  In all cases, except for the [NeV] 14.32 and [ClII]
14.36 lines, these uncertainties are the larger of either the
Gaussian fit or the difference in the two nod positions.  For
[NeV] 14.32 and [ClII] 14.36 the quoted errors are conservatively
estimated at $20\%$.  The FWHM reported in column five is from a
Gaussian fit to the SH or LH data, after removal of the
instrumental profile, in quadrature.  In all cases the error
given for the FWHM is the larger of either the fit, or the known
dispersion due to order curvature and undersampling ($\sim60$km
s$^{-1}$.  FWHM are not listed for unresolved emission lines measured only in the SL
data (those with wavelengths of $\lambda < 9.6\mu$m), or for the PAH features.  All lines were measured using the SMART
spectral reduction package (Higdon et al.  2004), except for the PAH 
features which were measured as a product of the 
multi-component model fits described in the text.} 

\end{deluxetable}

\end{document}